\documentclass[prd,showpacs,preprintnumbers,amsmath,amssymb]{revtex4}

\usepackage{graphicx}
\usepackage{dcolumn}
\usepackage{bm}

\def\be{\begin{eqnarray}}
\def\ee{\end{eqnarray}}
\def\bea{\begin{eqnarray}}
\def\eea{\end{eqnarray}}
\def\kT{{\bf k}_\perp}

\def\rT{{\bf r}_\perp}
\def\rp{{\bf r}_\perp}

\def\bp{{\bf b}_\perp}

\def\DT{{\bf \Delta}_\perp}
\def\Dp{{\bf \Delta}_\perp}
\def\Ds{{\bf \Delta}_\perp^2}
\def\0T{{\bf 0}_\perp}
\begin{document}


\title{GPDs with $\zeta \neq 0$}

\author{Matthias Burkardt}
 \affiliation{Department of Physics, New Mexico State University,
Las Cruces, NM 88003-0001, U.S.A.}
\date{\today}


\begin{abstract}
We revisit the light-cone wave function representation for 
generalized parton distributions with $\zeta\neq 0$. 
After translating the $t$-slope into a $\Delta_\perp^2$-slope, 
the two-dimensional Fourier transform of GPDs is interpreted
as the transition matrix element
as a function of the separation between the active quark and the
center of momentum of the spectators. In the limit 
$x\rightarrow \zeta$ it is discussed how
this information can be used to learn about the dependence of the
mean separation between the active quark and the spectators 
on the momentum fraction carried by the active quark.
\end{abstract}

\maketitle
\section{Introduction}
Hard exclusive processes, such as 
Deeply Virtual Compton Scattering (DVCS), $\gamma^*p\longrightarrow
\gamma p$, where $\gamma^*$ is a virtual photon with virtuality
$q^2=-Q^2<0$, have emerged as a novel probe for hadron structure.
In the Bjorken limit, and for a momentum transfer $t$ to the proton
that is much less than $Q^2$, the DVSC amplitude factorizes into a
convolution of the Compton amplitude off a quark, constituting the
hard part, and a quark correlation function, constituting
the soft part \cite{Ji,AR,Collins:1996fb}. 
The latter is parameterized by  Generalized
Parton Distributions GPDs, through their dependence on
the Bjorken variable $\zeta = \frac{Q^2}{2p\cdot q}$, and
the momentum fraction $x$ of the active quark before being
struck by the virtual photon. A physical interpretation for GPDs
is most easily available in the light-cone framework \cite{LCWFRep}
where, for $x>\zeta$ they represent the probability amplitude 
that the proton 
remains intact after a quark carrying momentum fraction $x$
absorbs a longitudinal momentum  $-\zeta$ (in units of the
initial proton momentum) and an invariant momentum transfer $t$. 
As in Deep-Inelastic Scattering (DIS), the variable $Q^2$ has the
interpretation of the spatial resolution. However, since GPDs
provide information about the distribution of partons in impact
parameter space \cite{Burkardt:2000za,Diehl:2002he,IJMPA}, the $Q^2$ dependence
in DVCS not only provides the scale dependence, but also the
`pixel-size' for the spatial images obtained from Fourier 
transforming
GPDs. Unfortunately, a probabilistic interpretation for the 
Fourier transforms of GPDs is restricted to $\zeta=0$
\cite{Burkardt:2001ni,Pobylitsa:2002iu}. Since the probabilistic
interpretation facilitates the development of intuitive models for
GPDs, most phenomenological models for GPDs are more reliable
for $\zeta=0$, and utilizing these models for $\zeta\neq 0$
gives rise to uncertainties that are difficult to quantify.
This is very unfortunate since DVCS  typically provides constraints
only for GPDs with $\zeta \neq 0$.
In particular, the imaginary part of the DVCS amplitude is only 
sensitive to $x=\zeta$
\be
\Im \left\{ T^{DVCS}\right\} \propto GPD(x=\zeta,\zeta,t,Q^2),
\ee
The real part appears in a convolution integral 
\be
\Re \left\{ T^{DVCS}\right\} \propto \int dx
\frac{GPD(x,\zeta,t,Q^2)}{x\pm\zeta},
\ee
where the factor $\frac{1}{x\pm\zeta}$ emphasizes the regions
$x\approx \pm\zeta$. 
In either case, understanding the vicinity of $x\approx \pm\zeta$
appears to be crucial for understanding the DVCS amplitude.
The goal of this note is to develop some intuition about 
GPDs in this important regime. More specifically, we will
consider the $t$-slopes of GPDs for $x\approx \zeta$ and what can
be learned from them.

\section{Light-Cone Wave Function Representation for GPDs}
Although the primary focus of this work is  
$x\approx \zeta$, we consider first $x>\zeta$, 
where GPDs are diagonal in Fock space. The regime $x=\zeta$ is 
then approached through a limiting procedure. For $x>\zeta$
simple overlap representations for GPDs in terms of
light-cone wave functions exist that resemble
overlap integrals for form factors in non-relativistic
systems \cite{LCWFRep,Diehl:2003ny}
\be
GPD(x,\zeta,t) &=& \sum_{n,\lambda_i}(1-\zeta)^{1-\frac{n}{2}}
\int \prod_{i=1}^n 
\frac{{\rm d}x_i {\rm d}{\bf k}_{\perp,i}}{16\pi^3} 
16\pi^3 \delta \left(1-\sum_{j=1}^n x_j\right) \delta \left(
\sum_{j=1}^n {\bf k}_{\perp j}\right)
\delta(x-x_1)\\
& & \quad \quad \quad \quad \times \psi^{s^\prime}_{(n)}(x_i^\prime, 
{\bf k}_{\perp i}^\prime, \lambda_i)^*
\psi^{s}_{(n)}(x_i, {\bf k}_{\perp i}, \lambda_i),
\label{eq:overlap}
\ee
where
\be
GPD(x,\zeta,t)= \frac{\sqrt{1-\zeta}}{1-\frac{\zeta}{2}}H(x,\zeta,t)
- \frac{\zeta^2}{4 \left(1-\frac{\zeta}{2}\right)\sqrt{1-\zeta}}
E(x,\zeta,t)
\ee
for $s^\prime=s$, and
\be
GPD(x,\zeta,t)=\frac{1}{\sqrt{1-\zeta}}\frac{\Delta^1-i\Delta^2}{2M}
E(x,\zeta,t)
\ee
for $s^\prime= \uparrow$ and $s= \downarrow$, and 
${\bf \Delta}$ is the transverse momentum transfer.
The arguments of the final state wave function are
$x_1^\prime = \frac{x_1-\zeta}{1-\zeta}$ and
${\bf k}_{\perp 1}^\prime ={\bf k}_{\perp 1} -\frac{1-x_1}{1-\zeta}
{\bf \Delta}_\perp$ for the active quark, and
$x_i^\prime = \frac{x_i}{1-\zeta}$ and
${\bf k}_{\perp 1}^\prime ={\bf k}_{\perp 1} +\frac{x_i}{1-\zeta}
{\bf \Delta}_\perp$ for the spectators $i=2,...,n$.

In order to elucidate the essential steps, we study first 
the simple case of a two-particle system (e.g. quark plus diquark),
where we consider light-cone wave functions as a function of
the distance between the active quark and the spectator
\be
\tilde{\psi}^s(x,\rT) = \int \frac{d^2\kT}{2\pi} \psi^s(x,\kT)
e^{i\kT\cdot \rT}.
\label{eq:FT}
\ee
Inserting the position space wave function (\ref{eq:FT})
diagonalizes the transverse part of the 
overlap integral in Eq. (\ref{eq:overlap}), yielding
\be
\int{d^2\kT} \psi^{s^\prime}(x^\prime,
{\bf k}_\perp^\prime)^*\psi^{s}(x,{\bf k}_\perp)
=
\int{d^2\rT} \tilde{\psi}^{s^\prime}(x^\prime,
{\bf r}_\perp)^*\tilde{\psi}^{s}(x,{\bf r}_\perp)
e^{-i\frac{1-x}{1-\zeta}\rT\cdot \DT}
\ee
For $\zeta \rightarrow 0$ one recovers the known result 
\cite{Burkardt:2000za}
that GPDs are Fourier transforms of the distribution
of partons in impact parameter space, where the impact parameter
$\bp= (1-x)\rT= {\bf r}_{\perp 1}-{\bf R}_\perp$ 
is the separation of the active quark
from the center of momentum 
${\bf R}_\perp \equiv x {\bf r}_{\perp 1}+ (1-x){\bf r}_{\perp 2}$.

For the general case, we also switch to transverse position
\be
\psi^{s}_{(n)}(x_i, {\bf k}_{\perp i}, \lambda_i) =
\int\prod_{i=1}^n \frac{d^2{\bf r}_{\perp i}}{2\pi}
e^{-i {\bf k}_{\perp i}\cdot {\bf r}_{\perp i}}
\tilde{\psi}^{s}_{(n)}(x_i, {\bf r}_{\perp i}, \lambda_i)
\label{eq:Fourier}.
\ee
Since we are dealing with plane wave states, one needs
to be careful with the normalization of these states and a more
careful treatment should involve working with wave packets. Here we 
will skip these tedious steps that have been studied carefully in 
Refs. \cite{Burkardt:2000za,Diehl:2002he} 
and immediately insert (\ref{eq:Fourier})
into (\ref{eq:overlap}), yielding
\be
GPD(x,\zeta,t) 
= \sum_{n,\lambda_i}(1-\zeta)^{1-\frac{n}{2}}
\int\prod_{i=1}^n \frac{d^2{\bf r}_{\perp i}}{2\pi}
\tilde{\psi}^{s^\prime}_{(n)}(x_i^\prime, 
{\bf r}_{\perp i}, \lambda_i)^*
\tilde{\psi}^{s}_{(n)}(x_i, {\bf r}_{\perp i}, \lambda_i)
e^{-\frac{i}{1-\zeta}\left({\bf r}_{\perp 1} - {\bf R}_\perp\right)
\cdot {\bf \Delta}_\perp}.
\ee
where ${\bf R}_\perp = \sum_i x_i{\bf r}_{\perp i}$ is the transverse
center of momentum of all partons in the initial state.

Since the transverse center of momentum changes in the process
\cite{Diehl:2002he},
it is useful to replace it by the separation between the
active quark and the center of momentum of the spectators
${\bf R}_{\perp s}$, using
\be
{\bf r}_\perp \equiv {\bf r}_{\perp 1} - {\bf R}_{\perp s}
= \frac{1}{1-x} \left({\bf r}_{\perp 1}-{\bf R}_\perp\right)
\ee
and one finds
\be
GPD(x,\zeta,t) 
= \sum_{n,\lambda_i}(1-\zeta)^{1-\frac{n}{2}}
\int\prod_{i=1}^n \frac{d^2{\bf r}_{\perp i}}{2\pi}
\tilde{\psi}^{s^\prime}_{(n)}(x_i^\prime, 
{\bf r}_{\perp i}, \lambda_i)^*
\tilde{\psi}^{s}_{(n)}(x_i, {\bf r}_{\perp i}, \lambda_i)
e^{-i\frac{1-x}{1-\zeta}\left({\bf r}_{\perp 1} - {\bf R}_{\perp s}\right)
\cdot {\bf \Delta}_\perp}.
\label{eq:convo2}
\ee

While GPDs for $x>\zeta> 0$ are still diagonal in the absolute
transverse
positions of all partons, they appear off-diagonal when positions
are measured relative to the $\perp$ center of momentum 
\cite{Diehl:2002he}.
However, as the momentum carried by the
active quark changes between initial and final state, 
so does the location
of the transverse center of momentum \cite{IJMPA}. Therefore,
even though the (absolute) $\perp$ positions of the active 
quark/spectators remain unchanged, their separation from the
$\perp$ center of momentum changes since the latter does.
For the physical interpretation of GPDs in the case of 
$\zeta \neq 0$, working with relative $\perp$ position coordinates 
(i.e. relative to each other) rather than impact parameter (measured
relative to the $\perp$ center of momentum may thus be preferable.
Indeed, the discussion above illustrates that, for nonzero $\zeta$, 
the Fourier transform of GPDs w.r.t. the transverse momentum
transfer $\DT$ yields information about the transition matrix element
between the initial and final state, when the $\perp$ distance
between the active quark and the center of momentum of the spectators
is ${\bf r}_\perp$.
More precisely, $\frac{1-\zeta}{1-x}{\bf r}_\perp$ is Fourier
conjugate to $\Dp$, and for $x=\zeta$, the variable conjugate to the
$\Dp$ is just ${\bf r}_\perp$.

\section{GPDs for $x\rightarrow \zeta$}
When $x=\zeta$, the coefficient multiplying 
$\rp \cdot \Dp$ in the exponent
in Eq. (\ref{eq:convo2}) becomes equal to one, i.e. in that limit
the Fourier transform of GPDs w.r.t. $\Dp$ yields the dependence
of the overlap matrix element on the separation $\rp$ between the
active quark and the center of momentum ${\bf R}_{\perp s}$
of the spectators.
While for $\zeta=0$ it is the separation from the center of momentum
of the whole hadron that sets the scale, it is the separation from
then center of momentum of the spectators that matters for 
$x=\zeta$. In order to utilise the above observations in the 
interpretation of GPDs, we note that \cite{LCWFRep}
\be
- t = \frac{\zeta^2 M^2+{\bf \Delta}_\perp^2}{1-\zeta}.
\label{eq:kinematical}
\ee
Therefore, if the $t$-dependence of GPDs is parameterized in the 
form
\be
GPD \propto e^{Bt}
\ee
one finds for the ${\bf \Delta}_\perp^2$-dependence 
\be
GPD \propto e^{-B_\perp {\bf \Delta}_\perp^2}
\ee
with $B_\perp = \frac{1}{1-\zeta}B$. 
Thus, even if the $\Ds$-slope (described by $B_\perp$) remains
finite as $\zeta \rightarrow 1$, the $t$-slope (described by $B$)
goes to zero.
This purely kinematical effect arises from the relation between
$t$ and $\Ds$ (\ref{eq:kinematical}) with 
$\zeta = 1-\frac{p^{+\prime}}{p^+}$ fixed. 

Since $\Dp$ is the momentum space variable conjugate
to $\rp={\bf r}_{\perp 1} -{\bf R}_{\perp s}$ (for $x=\zeta$), 
it is thus important to translate the $t$-dependence
of GPDs first into a $\Ds$-dependence before attempting to 
interpret the data. 

What should one thus expect for the $\zeta$-dependence of 
GPDs at $x=\zeta$? The relevant GPDs are proportional to the
the overlap between on initial state where the active quark 
carries momentum fraction $\zeta$ and a final state where the 
active quark carries almost no momentum. Intuitively one would
expect that the average separation between active 
quark and the spectators increases as the momentum fraction of the
active quark decreases, i.e. in this case the final state
wave function should be smaller than the initial state wave 
function.

In general, the overlap integral describing the GPDs 
(\ref{eq:overlap}) depends not only on the distribution of the
active quark but also on that of the spectators. However, it appears
reasonable to assume that the spectator wave function (for a given
position of the spectator center of momentum) does not
depend very strongly on the position of the active quark when
the active quark is far away from the spectators.
In the following we will thus make the simplifying assumption that
the overlap integral for the spectators (at fixed $x$ and $\zeta$)
does not depend on the separation of the active quark from the
spectators. This does not mean that the spectators wave function is 
point-like!

In order to qualitatively understand how the above overlap integrals
depend on $\zeta$ (for , we rescale all momentum fractions in units of
the final state momentum, i.e. the initial state hadron
carries momentum $1/(1-\zeta)$ and the final state hadron carries
momentum $1$. As the active quark carries momentum fraction $0$,
nothing in the final state depends on $\zeta$ and hence the
$\zeta$-dependence arises from the change in the initial state
wave function, and the resulting change in the overlap integrals.
This observation suggests the following interpretation for
the $\zeta$ dependence of the $\Delta_\perp^2$-slope of GPDs.
For instance, if the $\Delta_\perp^2$-slope decreases with 
increasing
$\zeta$, that would be an indication that the mean separation 
between the active quark and the spectators decreases with the
momentum fraction carried by the active quark.

If one neglects the $\Delta_\perp^2$-dependence of the overlap 
integral for the spectators, one can use this reasoning to extract
the `size' (mean separation of the active quarks from the
spectators) as a function of the momentum fraction carried by the
active quark. For example, when the initial and final
state wave function are proportional to $e^{-\frac{r_\perp}{R_1}}$
and $e^{-\frac{r_\perp}{R_2}}$ then the effective radius appearing
in the product is the harmonic mean of the rms radii of the
individual wave functions squared
$\frac{1}{R_{eff}} = 
\frac{1}{2}\left(\frac{1}{R_{1}}+\frac{1}{R_{2}}\right)$.

\section{Summary}
For $\zeta\neq0$, the two dimensional Fourier transform of
GPDs is more easily interpretable if one introduces the
separation ${\bf r}_\perp$
between the active quark and the center of momentum
of the spectators, as this variable is the same in the initial
and final state of the hadronic matrix element defining the
GPDs. The ${\bf r}_\perp$ 
dependence of the matrix element is obtained by
Fourier transforming GPDs with a factor 
$e^{-i\frac{1-x}{1-\zeta}{\bf r}\cdot {\bf \Delta}_\perp}$, i.e.
for $x=\zeta$ the variable ${\bf r}$ is Fourier conjugate to
${\bf \Delta}_\perp$.

The mean ${\bf r}^2_\perp$, and hence the $\Delta_\perp^2$-slope
of GPDs should be a typical hadronic scale. Therefore the
$t$-slope, which is related to the $\Delta_\perp^2$-slope by a
kinematic factor of $1-\zeta$, should go to
zero as $\zeta\rightarrow 1$, even if the wave function does
not become point-like. The $t$-slope divided by $1-\zeta$ can be 
used to study how the mean separation of the active quark
from the center of momentum of the spectators varies with
$\zeta$. Intuitively, one would expect this `size' to decrease
with $\zeta$. Application of the above procedure to deeply-virtual
meson production indeed yields a size that decreases with increasing 
$\zeta$ \cite{stoler}. 
DVCS data for the $t$-slope  \cite{CLAS} also shows a decrease
with increasing $\zeta$.

{\bf Acknowledgements:}
I would like to thank V. Burkert and V. Kubarovsky for useful 
discussions.
This work was supported by the DOE under grant number 
DE-FG03-95ER40965.

\bibliography{xi.bbl}
\end{document}